# Radiation Reaction in Quantum Vacuum

Keita Seto

*ELI-NP, "Horia Hulubei" National Institute of Physics and Nuclear Engineering,*

*Str. Atomistilor, nr.407, localitatea Magurele, jud. Ilfov, 077125, Romania*

*keita.seto@eli-np.ro

...........................................................

After the development of the radiating electron theory by P. A. M. Dirac in 1938, many authors have tried to reformulate this model named "radiation reaction". Recently, this equation has become important for ultra-intense laser-electron (plasma) interactions. In our recent research, we found a stabilized model of radiation reaction in quantum vacuum [PTEP 2014, 043A01 (2014).]. It led us to an updated Fletcher-Millikan's charge to mass ratio including radiation. In this paper, I will discuss the generalization of our previous model and the new equation of motion with radiation reaction in quantum vacuum via photon-photon scatterings and also introduce the new tensor $d\mathcal{E}/dm$, as the anisotropy of the charge to mass ratio.

...........................................................

## 0. Definitions

I defined the Minkowski spacetime as the mathematical set of $(\mathbb{A}^4, g)$. $\mathbb{A}^4$ is a 4-dimentional affine space and $g$ is the Minkowski metric with the signature of $(+,-,-,-)$. In the definition of any affine space, it has a sub-vector space as the structure of the affine space. Now, let's define that model-linear space $\mathbb{V}_M^4$ as the structure of $\mathbb{A}^4$. In this paper, the 4-dimensional contravariant vector space is denoted $\mathbb{V}_M^4$ which has the basis of $\{\mathbf{e}_{\mu=0,1,2,3}\} \in \mathbb{V}_M^4$, and the 4-dimensional covariant vector space is $*\mathbb{V}_M^4$ with the basis of $\{\boldsymbol{\omega}_{\mu=0,1,2,3}\}$ as the dual space of $\mathbb{V}_M^4$. The metric is defied by $g = g_{\alpha\beta}\boldsymbol{\omega}^\alpha \otimes \boldsymbol{\omega}^\beta \in *\mathbb{V}_M^4 \otimes *\mathbb{V}_M^4$. Following this, Let's $\forall w \in \mathbb{V}_M^4$, the covariant vector of it is,

$$w^\flat = g(\bullet, w) = g_{\alpha\beta}w^\beta \boldsymbol{\omega}^\alpha \in *\mathbb{V}_M^4. \tag{0-1}$$

For $\forall A, B \in \mathbb{V}_M^4 \otimes \mathbb{V}_M^4$, I defied the map $\langle \bullet | \circ \rangle : (\mathbb{V}_M^4 \otimes \mathbb{V}_M^4) \times (\mathbb{V}_M^4 \otimes \mathbb{V}_M^4) \to \mathbb{R}$,

$$\langle A | B \rangle = A_{\mu\nu} B^{\mu\nu} \in \mathbb{R} \tag{0-2}$$

$F \in \mathbb{V}_M^4 \otimes \mathbb{V}_M^4$ is considered as the map $*\mathbb{V}_M^4 \to \mathbb{V}_M^4$,

$$F[\bullet, w^\flat] = F^{\mu\nu} w_\nu \mathbf{e}_\mu \in \mathbb{V}_M^4. \tag{0-3}$$

With $\forall \mathcal{K} \in \mathbb{V}_M^4 \otimes \mathbb{V}_M^4 \otimes \mathbb{V}_M^4 \otimes \mathbb{V}_M^4$,

$$\mathcal{K}[\bullet, \circ, F^\flat] = \mathcal{K}^{\mu\nu\alpha\beta} F_{\alpha\beta} \mathbf{e}_\mu \otimes \mathbf{e}_\nu \in \mathbb{V}_M^4 \otimes \mathbb{V}_M^4 \tag{0-4}$$

$$\mathcal{K}[\bullet, w^\flat, A^\flat] = \mathcal{K}^{\mu\nu\alpha\beta} w_\nu F_{\alpha\beta} \mathbf{e}_\mu \in \mathbb{V}_M^4 \tag{0-5}$$

The dual tensor of $F$ is

$$*F = *F^{\alpha\beta}(\mathbf{e}_\alpha \otimes \mathbf{e}_\beta) = \frac{\varepsilon^{\alpha\beta\mu\nu} F_{\mu\nu}}{2!}(\mathbf{e}_\alpha \otimes \mathbf{e}_\beta) = 1/2! \times \varepsilon[\bullet, \circ, F^\flat] \tag{0-4}$$

Here, $F^\flat = F_{\mu\nu} \omega^\mu \otimes \omega^\nu \in *\mathbb{V}_M^4 \otimes *\mathbb{V}_M^4$ and $\varepsilon \in \mathbb{V}_M^4 \otimes \mathbb{V}_M^4 \otimes \mathbb{V}_M^4 \otimes \mathbb{V}_M^4$ is the Levi-Civita's tensor.

## 1. Introduction

In 1938, P. A. M. Dirac proposed the equation of an electron's motion in classical-relativistic dynamics including the electron's self-interaction, the so-called the Lorentz-Abraham-Dirac (LAD) equation [1].

$$m_0 \frac{dw}{d\tau} = -eF_{ex}[\bullet, w^\flat] - eF_{LAD}[\bullet, w^\flat] \in \mathbb{V}_M^4 \tag{1}$$

Here, $m_0, e$ and $\tau \in \mathbb{R}$ are the rest mass, the charge and the proper time of an electron. $w \in \mathbb{V}_M^4$ is the 4-velocity defined by $w = \gamma(c, \mathbf{v})$. $F_{ex} \in \mathbb{V}_M^4 \otimes \mathbb{V}_M^4$ is an arbitrary external field. The field $F_{LAD} \in \mathbb{V}_M^4 \otimes \mathbb{V}_M^4$ is the reaction field, which acts on the electron, feeding back on the electron due to light emission. This field is defined by using the retarded field $F_{ret}$ and the advanced field $F_{adv}$,

$$F_{LAD}\big|_{x=x(\tau)} = \frac{F_{ret} - F_{adv}}{2}\bigg|_{x=x(\tau)} = -\frac{m_0 \tau_0}{ec^2}\left(\frac{d^2w}{d\tau^2} \otimes w - w \otimes \frac{d^2w}{d\tau^2}\right) \in \mathbb{V}_M^4 \otimes \mathbb{V}_M^4. \tag{2}$$

The constant $\tau_0$ is $\tau_0 = e^2/6\pi\varepsilon_0 m_0 c^3 = O(10^{-24})$. Following the above considerations, he upgraded to the relativistic force equation from the non-relativistic equation of H. A. Lorentz [2] and M. Abraham [3].

$$f_{LAD} = -eF_{LAD}[\bullet, w^\flat] = m_0 \tau_0 \frac{d^2w}{d\tau^2} + \frac{m_0 \tau_0}{c^2} g\left(\frac{dw}{d\tau}, \frac{dw}{d\tau}\right) w \in \mathbb{V}_M^4 \tag{3}$$

This is called the LAD radiation reaction force. J. Schwinger derived the Larmor's formula

$$\frac{dW}{dt} = -m_0 \tau_0 g\left(\frac{dw}{d\tau}, \frac{dw}{d\tau}\right) = m_0 c^2 \tau_0 \frac{\dot{\boldsymbol{\beta}}^2 - (\boldsymbol{\beta} \times \dot{\boldsymbol{\beta}})^2}{(1-\boldsymbol{\beta}^2)^3} \tag{4}$$

by using this LAD field $F_{LAD}$ [3]. We can find this Larmor's formula as the coefficient in Eq.(3). The second term on the R. H. S. of Eq.(3) is the so-called "direct radiation term", therefore, this LAD equation has been treated as the equation of an electron's motion with light emission. For this reason, the LAD equation is a standard model of a radiating electron under ultra-high intense lasers. With the rapid progress of ultra-short pulse laser technology, the maximum intensities of these lasers has reached the order of $10^{22}$W/cm$^2$ [5, 6]. One laser facility, which can achieve such ultra-high intensity is LFEX (Laser for fast ignition experiment) at the Institute of Laser Engineering (ILE), Osaka University [7] and another is the next laser generation project, the Extreme Light Infrastructure (ELI) project [8] in Europe. If the laser intensity is higher than $10^{22}$W/cm$^2$, strong bremsstrahlung might occur. Accompanying this, the radiation reaction force (or damping force) can have a strong influence on the charged particle [9]. This is in spite of the fact that the LAD equation has a very significant mathematical problem. The solution of the LAD equation has the factor of an exponential. Let the vector function be $f \in \mathbb{V}_M^4$, the the solution of the LAD equation is

$$\frac{dw}{d\tau}(\tau) = f(\tau) \times \exp\frac{\tau}{\tau_0}. \tag{5}$$

This solution is derived by integration of the LAD equation, but this solution goes rapidly to infinity, since $\tau_0 = O(10^{-24})$ is a very small value [10, 11]. We call this run-away depending on the first term in Eq.(3) named the Schott term. It should be avoided to solve the equation stably.

For avoidance of this run-away problem, we've considered a radiating electron with a dress of field, in our previous paper [12].

$$\frac{d}{d\tau}w = -\frac{e}{m_0(1-\eta\langle F_{LAD}|F_{LAD}\rangle)}(F_{ex} + F_{LAD})[\bullet, w^\flat] \tag{6}$$

I named here this equation the Seto-Zhang-Koga (SZK) equation for instance. This dressed electron was described by vacuum polarization via the Heisenberg-Euler Lagrangian density [13, 14]. The dress stabilizes run-away by changing the coupling constant $e/m_0 \times (1-\eta\langle F_{LAD}|F_{LAD}\rangle)^{-1}$. However, the previous model considered only the correction of radiation from an electron. Moreover, the introduction of the external field was artificial (Eq.(24) in [12]).

To address these points, I will introduce a new model of radiation reaction which incorporates a smooth installation of the external fields, including the radiation-external field interaction in this paper. To achieve this, we first consider a more general equation of motion with radiation reaction in quantum vacuum in Section 2. But, we will not investigate a more concrete dynamics of quantum vacuum beyond the Heisenberg-Euler's vacuum. In this phase, we only assume the Lagrangian density is a function of $\langle F|F\rangle$ and $\langle F|*F\rangle$. Next, I will proceed to a concrete model by using the lowest order Heisenberg-Euler Lagrangian density as the model of quantum vacuum in Section 3. I will present the stability of the new equation via analysis and numerical calculations. Finally, this will lead us to an anisotropic correction for the charge to mass ratio by R. Fletcher and H. Millikan [15, 16].

**2. Derivation of a new method of radiation reaction**

The Heisenberg-Euler Lagrangian density includes the dynamics of the quantum vacuum correction. However this is only suitable for constant fields. In this chapter, let's consider the general Lagrangian density for quantum vacuum without a concrete definition. It should be described by $\langle F|F\rangle$ and $\langle F|{*F}\rangle$ like the Heisenberg-Euler Lagrangian density. Here, $F \in \mathbb{V}_M^4 \otimes \mathbb{V}_M^4$ is the electromagnetic tensor and $*F$ is the dual tensor of $F$. Now, the Lagrangian density for propagating photons is,

$$L(\langle F|F\rangle, \langle F|{*F}\rangle) = -\frac{1}{4\mu_0}\langle F|F\rangle + L_{\text{Quantum Vacuum}}(\langle F|F\rangle, \langle F|{*F}\rangle). \qquad (7)$$

Of course, this Lagrangian density $L_{\text{Quantum Vacuum}}$ needs to converge to the Heisenberg-Euler Lagrangian density when the field $F$ is a constant field. For instance, we assume that $L$ and $L_{\text{Quantum Vacuum}}$ are functions of $C^\infty$. From this equation, the Maxwell equation is derived as follows:

$$\partial_\mu \left[ F^{\mu\nu} - \eta f \times F^{\mu\nu} - \eta g \times {*F}^{\mu\nu} \right] = 0 \qquad (8)$$

$$\eta f(\langle F|F\rangle, \langle F|{*F}\rangle) = 4\mu_0 \frac{\partial L_{\text{Quantum Vacuum}}}{\partial \langle F|F\rangle} \qquad (9)$$

$$\eta g(\langle F|F\rangle, \langle F|{*F}\rangle) = 4\mu_0 \frac{\partial L_{\text{Quantum Vacuum}}}{\partial \langle F|F\rangle} \qquad (10)$$

In these equations, $\eta = 4\alpha^2 \hbar^3 \varepsilon_0 / 45 m_0^4 c^3$. The field

$$\frac{1}{c\varepsilon_0} M = -\eta f \times F - \eta g \times {*F} \qquad (11)$$

represents the vacuum "polarization", therefore, $F - \eta f \times F - \eta g \times {*F}$ refers to the dressed field set of $(\mathbf{D}, \mathbf{H})$. In addition, the following is satisfied $\partial_\mu (F_{\text{ex}}{}^{\mu\nu} + F_{\text{LAD}}{}^{\mu\nu}) = 0$. Thus, Eq.(8) suggests the connection between $F - \eta f \times F - \eta g \times {*F}$ and $(\mathbf{D}, \mathbf{H}) = F_{\text{ex}} + F_{\text{LAD}}$ with the continuity and smoothness with $C^\infty$ on all points in the Minkowski spacetime. At a point far from an electron, the external fields are given and radiation can be observable [Fig.1]. At this point, Eq.(8) becomes,

$$\boxed{F - \eta f \times F - \eta g \times {*F} = \mathfrak{F}}. \qquad (12)$$

Here, it is denoted as $\mathfrak{F} = F_{\text{ex}} + F_{\text{LAD}} \in \mathbb{V}_M^4 \otimes \mathbb{V}_M^4$ for instance. In our previous model [12], we assumed

$$F - \eta f \times F - \eta g \times {*F} = F_{\text{LAD}}. \qquad (13)$$

Therefore, we didn't consider the correction of the external field. I could incorporate the external field naturally here. This is the most important difference between the new and old model.

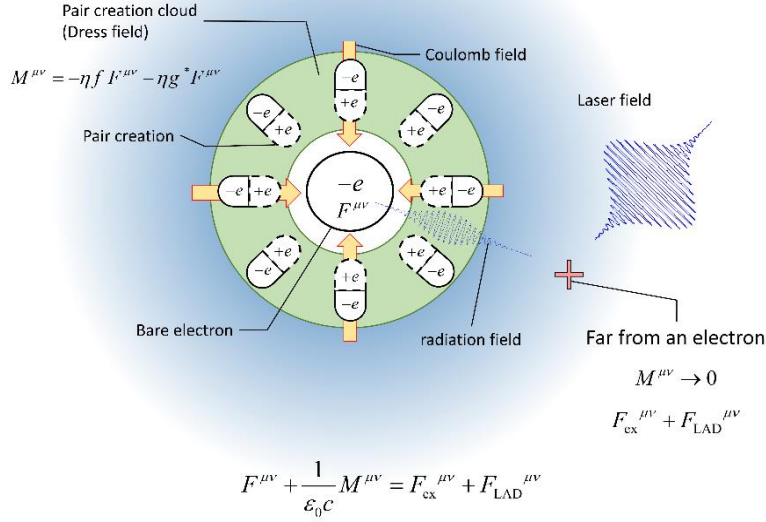

Fig. 1    The bare field and dressed field.

By using the field's continuity and smoothness, Eq.(12) can be applied to, not only points far from an electron, but at the electron point. Our interest is the bare (undressed) field $F = (\mathbf{E}, \mathbf{B})$ at the point of an electron for defining the electromagnetic force $-eF[\bullet, w^{\flat}] = -eF^{\mu\nu}w_\nu \mathbf{e}_\mu$. We consider the description of the tensor $F$ from Eq.(12) as the way to obtain the solution.

$$\mathcal{L}^{\mu\nu\alpha\beta} F_{\alpha\beta} = \mathfrak{F}^{\mu\nu} \tag{14}$$

$$\mathcal{L}^{\mu\nu\alpha\beta} = g^{\mu\alpha} g^{\nu\beta} (1-\eta f) - \frac{1}{2!}\varepsilon^{\mu\nu\alpha\beta} \times \eta g \tag{15}$$

Here, $\mathcal{L}$ is the permittivity tensor in Minkowski spacetime. However, we define a new tensor,

$$\begin{aligned}\bar{\mathcal{K}}_{\rho\sigma\mu\nu} &= \frac{g_{\rho\mu}g_{\sigma\nu}(1-\eta f) + \frac{1}{2!}\varepsilon_{\rho\sigma\mu\nu} \times \eta g}{(1-\eta f)^2 + (\eta g)^2} \\ &= \frac{1}{1-\eta f} \times \frac{1}{1+\frac{(\eta g)^2}{(1-\eta f)^2}} \left( g_{\rho\mu}g_{\sigma\nu} + \frac{1}{2!}\varepsilon_{\rho\sigma\mu\nu} \times \frac{\eta g}{1-\eta f} \right).\end{aligned} \tag{16}$$

From the relation in which $\varepsilon_{\rho\sigma\mu\nu}\varepsilon^{\mu\nu\alpha\beta} = -2(\delta_\rho{}^\alpha \delta_\sigma{}^\beta - \delta_\rho{}^\beta \delta_\sigma{}^\alpha)$ and considering the anti-symmetry of $F$, $\bar{\mathcal{K}}_{\rho\sigma\mu\nu}\mathcal{L}^{\mu\nu\alpha\beta} F_{\alpha\beta} = F_{\rho\sigma}$. Therefore, the field $F$ becomes,

$$F^{\mu\nu} = \bar{\mathcal{K}}^{\mu\nu\rho\sigma}\mathfrak{F}_{\rho\sigma} = \frac{1}{1-\eta f} \times \frac{1}{1+\frac{(\eta g)^2}{(1-\eta f)^2}}\left[\mathfrak{F}^{\mu\nu} + \frac{\eta g}{1-\eta f}\times{}^*\mathfrak{F}^{\mu\nu}\right]. \quad (17)$$

Since the style of the equation of motion is,

$$m_0 \frac{dw}{d\tau} = -eF[\bullet, w^\flat] \quad \in \mathbb{V}_M^{\ 4} \quad (18)$$

by substitution of Eq.(17) into Eq.(18), we obtain,

$$m_0(1-\eta f)\left[1+\frac{(\eta g)^2}{(1-\eta f)^2}\right]\frac{dw}{d\tau} = -e\mathfrak{F}[\bullet, w^\flat] - e\frac{\eta g}{1-\eta f}{}^*\mathfrak{F}[\bullet, w^\flat]$$

$$\Rightarrow \quad m_0(1-\eta f_0)\frac{dw}{d\tau} = -e\mathfrak{F}[\bullet, w^\flat] - e\eta g_0{}^*\mathfrak{F}[\bullet, w^\flat] + O(\eta^2). \quad (19)$$

Here, I used the Taylor's expansion of $f$ and $g$ near $\eta = 0$. Paying attention to the relation $F|_{\eta=0} = \mathfrak{F}$ and denoting that $f_0 = f(\langle\mathfrak{F}|\mathfrak{F}\rangle, \langle\mathfrak{F}|{}^*\mathfrak{F}\rangle)$ and $g_0 = g(\langle\mathfrak{F}|\mathfrak{F}\rangle, \langle\mathfrak{F}|{}^*\mathfrak{F}\rangle)$ for the simplification,

$$f = f_0 + \eta\delta f + O(\eta^2), \quad (20)$$

$$g = g_0 + \eta\delta g + O(\eta^2). \quad (21)$$

By treating the first order of quantum vacuum,

$$m_0 \frac{dw}{d\tau} = -e\frac{\mathfrak{F}[\bullet, w^\flat] + \eta g_0{}^*\mathfrak{F}[\bullet, w^\flat]}{1-\eta f_0}. \quad (22)$$

Where, introducing a new tensor $\mathcal{K}$ like Eq.(16) with $G = g^{\mu\alpha}g^{\nu\beta}\partial_\mu \otimes \partial_\nu \otimes \partial_\alpha \otimes \partial_\beta$,

$$\mathcal{K} = \frac{G + \eta g_0 \times \frac{1}{2!}\varepsilon}{1-\eta f_0}, \quad (23)$$

the field is modified as

$$F = \mathcal{K}[\bullet, \circ, F^\flat] = \mathcal{K}^{\mu\nu\alpha\beta}\mathfrak{F}_{\alpha\beta}\mathbf{e}_\mu \otimes \mathbf{e}_\nu \in \mathbb{V}_M^{\ 4} \otimes \mathbb{V}_M^{\ 4}. \quad (24)$$

Finally, we need to pay attention to the fact that Eq. (24) is already included the radiation reaction field and quantum vacuum effects via the definition of Eq. (12). By rewriting Eq.(22),

$$\boxed{m_0 \frac{dw}{d\tau} = -e\mathcal{K}[\bullet, w^\flat, \mathfrak{F}^\flat] \quad \in \mathbb{V}_M^{\ 4}}. \quad (25)$$

This is the general formula of radiation reaction in quantum vacuum. The limit of $\hbar \to 0$ leads to a smooth connection to the LAD equation, since $\eta = 4\alpha^2\hbar^3\varepsilon_0/45m_0^4 c^3$ and $\mathcal{K} \to G$.

## 3. First order Heisenberg-Euler quantum vacuum

### 3.1 Equation of motion

In Section 2, the quantum vacuum was assumed to be functions of $\langle F | F \rangle$ and $\langle F | *F \rangle$ without concrete formulations. The Heisenberg-Euler Lagrangian density expresses the dynamics of quantum vacuum, but can be applied only for constant fields. However, its lowest order should be contained in $L_{\text{Quantum Vacuum}}$ [12]. Therefore, in this chapter, I assume that,

$$L_{\text{Quantum Vacuum}} = L_{\substack{\text{the lowest order of}\\\text{Heisenberg-Euler}}} = \frac{\alpha^2 \hbar^3 \varepsilon_0^2}{360 m_0^4 c} \left[ 4\langle F | F \rangle^2 + 7\langle F | *F \rangle^2 \right]. \tag{26}$$

In this case, instead of Eq. (12),

$$F - \eta \langle F | F \rangle \times F - \frac{7}{4} \eta \langle F | *F \rangle \times *F = \mathfrak{F}, \tag{27}$$

by using perturbations, $f_0$ and $g_0$ are

$$f_0 = \langle \mathfrak{F} | \mathfrak{F} \rangle = \langle F_{\text{LAD}} | F_{\text{LAD}} \rangle + 2\langle F_{\text{LAD}} | F_{\text{ex}} \rangle \tag{28}$$

$$g_0 = \frac{7}{4} \langle \mathfrak{F} | *\mathfrak{F} \rangle = \frac{7}{2} \langle F_{\text{LAD}} | *F_{\text{ex}} \rangle \tag{29}$$

Here, I used the relation that $\partial_\mu F_{\text{ex}}^{\mu\nu} = 0 \Rightarrow \langle F_{\text{ex}} | F_{\text{ex}} \rangle = 0$, $\langle F_{\text{ex}} | *F_{\text{ex}} \rangle = 0$ and $\langle F_{\text{LAD}} | *F_{\text{LAD}} \rangle \equiv 0$ [12]. Eq.(24) becomes,

$$F = \mathcal{K}[\bullet, \circ, \mathfrak{F}^\flat] = \frac{1}{1 - \eta f_0} \mathfrak{F} + \frac{\eta g_0}{1 - \eta f_0} *\mathfrak{F} \in \mathbb{V}_M^4 \otimes \mathbb{V}_M^4. \tag{30}$$

When $1 - \eta f_0 = 0$, the field $F$ becomes infinity and run-away appears. It is required that $1 - \eta f_0 > 0$ for application. From the relations, $\langle F_{\text{LAD}} | F_{\text{LAD}} \rangle = 2/e^2 c^2 \times g(f_{\text{LAD}}, f_{\text{LAD}}) = -2(m_0 \tau_0/ec)^2 \ddot{\mathbf{v}}^2 |_{\text{rest}} \leq 0$ and $\langle F_{\text{LAD}} | F_{\text{ex}} \rangle = 2m_0 \tau_0/ec^2 \times \ddot{\mathbf{v}} \cdot \mathbf{E}_{\text{ex}} |_{\text{rest}}$ in an electron's rest frame,

$$1 - \eta f_0 = \frac{2\eta}{ec} \times (m_0 \tau_0 \ddot{\mathbf{v}} |_{\text{rest}} - e\mathbf{E}_{\text{ex}} |_{\text{rest}})^2 + 1 - \frac{2\eta \mathbf{E}_{\text{ex}}^2 |_{\text{rest}}}{c^2} > 1 - \frac{2\eta \mathbf{E}_{\text{ex}}^2 |_{\text{rest}}}{c^2} \overset{\substack{\text{physical}\\\text{requirements}}}{>} 0. \tag{31}$$

The stability only depends on the external field in the rest frame of an electron. By using the Schwinger limit field $E_{\text{Schwinger}} = m_0^2 c^3 / e\hbar$, $1 - \eta f_0 > 1 - (5.2 \times 10^{-5}) \times (\mathbf{E}_{\text{ex}} |_{\text{rest}} / E_{\text{Schwinger}})^2$. The field $\mathbf{E}_{\text{ex}} |_{\text{rest}}$ should be treated below the Schwinger limit, therefore, $|\mathbf{E}_{\text{ex}}| \ll E_{\text{Schwinger}}$ is normally satisfied. Therefore, we require choices which satisfy Eq.(31) for $1 - \eta f_0 > 0$. Now, the dependence of $\langle F_{\text{LAD}} | *F_{\text{ex}} \rangle = 2m_0 \tau_0/ec \times \ddot{\mathbf{v}} \cdot \mathbf{B}_{\text{ex}} |_{\text{rest}}$ or $g_0$ are unsolved. If the external fields are absent, this field converges to our previous model [12].

$$F|_{F_{\text{ex}}=0} = \frac{1}{1 - \eta \langle F_{\text{LAD}} | F_{\text{LAD}} \rangle} F_{\text{LAD}} \tag{32}$$

Therefore, this new model is the generalization of our previous model. It can adopt quantum vacuum not only via radiation reaction, but via external fields like lasers. The equation of motion is

$$\frac{dw}{d\tau} = -\frac{e}{m_0(1-\eta\langle\mathfrak{F}|\mathfrak{F}\rangle)}\left(\mathfrak{F}[\bullet, w^\flat] + \frac{7}{4}\eta\langle\mathfrak{F}|*\mathfrak{F}\rangle *\mathfrak{F}[\bullet, w^\flat]\right). \tag{33}$$

### 3.2 Run-away avoidance

My previous model could avoid run-away (the effect of the self-acceleration) [12]. In this section, I will show that my new equation can avoid run-away by using a two-stage analysis. The first is the investigation of the radiation upper limit and the second is the asymptotic analysis proposed by F. Röhrich [17]. The run-away solution deeply depends on the infinite radiation from an electron. The physical meaning of run-away is a time-continuously infinite light emission via stimulations by an electron's self-radiation. In another words, when we can limit the value of the radiation, we can say the model avoids run-away. To check the stability of this equation, we consider the equation as follows derived from Eq.(33).

$$g\left(\frac{dw}{d\tau},\frac{dw}{d\tau}\right) = \frac{1}{m_0^2}\frac{\frac{e^2c^2}{2\eta}\eta f_0 + \frac{2}{7\eta}e^2c^2(\eta g_0)^2}{(1-\eta f_0)^2} + \frac{1}{m_0^2}\frac{g(f_{\text{ex}}+\eta g_0*f_{\text{ex}}, f_{\text{ex}}+\eta g_0*f_{\text{ex}})}{(1-\eta f_0)^2} \tag{34}$$

Here, I defined the forces $f_{\text{ex}} = -eF_{\text{ex}}^{\mu\nu}w_\nu\mathbf{e}_\mu$ and $*f_{\text{ex}} = -e(*F_{\text{ex}})^{\mu\nu}w_\nu\mathbf{e}_\mu \in \mathbb{V}_M^4$. In the rest frame, $f_0 = -2(m_0\tau_0/ec)^2\ddot{\mathbf{v}}^2|_{\text{rest}} + 4m_0\tau_0/ec^2 \times \ddot{\mathbf{v}}\cdot\mathbf{E}_{\text{ex}}|_{\text{rest}} = O(\ddot{\mathbf{v}}_{\text{rest}}^2)$ and $g_0 = 7m_0\tau_0/ec \times \ddot{\mathbf{v}}\cdot\mathbf{B}_{\text{ex}}|_{\text{rest}} = O(\ddot{\mathbf{v}}_{\text{rest}})$ are satisfied. When we face the run-away solution, then $|\ddot{\mathbf{v}}_{\text{rest}}| \to \infty$. Therefore, $O(|g_0|) < O(|f_0|)$ in the run-away case. Under the condition of Eq.(31),

$$\left|g\left(\frac{dw}{d\tau},\frac{dw}{d\tau}\right)\right|$$

$$\leq \frac{1}{m_0^2}\frac{e^2c^2}{2\eta}\frac{|\eta f_0|}{|1-\eta f_0|^2} + \frac{1}{m_0^2}\frac{2e^2c^2}{7\eta}\frac{|\eta g_0|^2}{|1-\eta f_0|^2}$$

$$+ \frac{|g(f_{\text{ex}},f_{\text{ex}})|}{m_0^2}\frac{1}{|1-\eta f_0|^2} + 2\frac{|g(f_{\text{ex}},*f_{\text{ex}})|}{m_0^2}\frac{|\eta g_0|}{|1-\eta f_0|^2} + \frac{|g(*f_{\text{ex}},*f_{\text{ex}})|}{m_0^2}\frac{|\eta g_0|^2}{|1-\eta f_0|^2}$$

$$\stackrel{\text{run-away}}{<} \frac{1}{m_0^2}\frac{e^2c^2}{2\eta}\frac{|\eta f_0|}{|1-\eta f_0|^2} + \frac{1}{m_0^2}\frac{2e^2c^2}{7\eta}\frac{|\eta f_0|^2}{|1-\eta f_0|^2}$$

$$+ \frac{|g(f_{\text{ex}},f_{\text{ex}})|}{m_0^2}\frac{1}{|1-\eta f_0|^2} + 2\frac{|g(f_{\text{ex}},*f_{\text{ex}})|}{m_0^2}\frac{|\eta f_0|}{|1-\eta f_0|^2} + \frac{|g(*f_{\text{ex}},*f_{\text{ex}})|}{m_0^2}\frac{|\eta f_0|^2}{|1-\eta f_0|^2}$$

$$< \infty \tag{35}$$

since the functions of $1/|1-x|^2$, $|x|/|1-x|^2$ and $|x|^2/|1-x|^2$ are finite in the domain $x \in (-\infty, 1)$. Now, $x = \eta f_0 \leq 2\eta\mathbf{E}_{\text{ex}}^2|_{\text{rest}}/c^2 < 1$ from Eq.(31). When we are in the case of run away, $dw/d\tau$ also becomes infinite because it is the integral of $d^2w/d\tau^2$, then $|g(dw/d\tau, dw/d\tau)| \to \infty$. But, this conflicts with the inequality of Eq.(35). Therefore, the Larmor formula becomes

$$\frac{dW}{dt} = -m_0\tau_0 g\left(\frac{dw}{d\tau},\frac{dw}{d\tau}\right) < \infty \tag{36}$$

for the whole time domain and run-away doesn't appear. Under the external field condition Eq.(31), solving the Eq.(33),

$$\frac{dw}{d\tau}(\tau) = \frac{e^{\frac{\tau}{\tau_0}}}{m_0\tau_0}\int_\tau^\infty d\tau' \left[ f_{\text{ex}} + \eta g_0 * f_{\text{ex}} + \frac{m_0\tau_0}{c^2} g\left(\frac{dw}{d\tau},\frac{dw}{d\tau}\right)w \right] \times e^{-\frac{\tau'}{\tau_0}} \times e^{\int_\tau^{\tau'}\frac{d\tau''}{\tau_0}\eta f_0}. \tag{37}$$

If we choose $\eta = 0$, this solution becomes Eq.(I) in Röhrlich's article [17]. He derived the "asymptotic" boundary condition in $\tau \to \infty$ by using l'Hôpital's rule. This is a method based on the our normal idea, "when $f_{\text{ex}} \in \mathbb{V}_M^4$ vanishes in $\tau \to \infty$, $dw/d\tau(\infty) \in \mathbb{V}_M^4$ also vanishes". He suggested that when run-away exists, then $dw/d\tau$ is not zero because of the self-stimulation by radiation. Therefore, $dw/d\tau(\infty) = 0$ is required for the model stability. We apply this distinction for Eq.(37). From l'Hôpital's rule, Eq.(37) becomes,

$$m_0\frac{dw}{d\tau}(\infty) = f_{\text{ex}}(\infty) + \eta g_0 * f_{\text{ex}}(\infty) + \frac{m_0\tau_0}{c^2} g\left(\frac{dw}{d\tau},\frac{dw}{d\tau}\right)w(\infty). \tag{38}$$

Here, I used the signature of the limit by Röhlrich. When the given $f_{\text{ex}}(\infty)$ and $*f_{\text{ex}}(\infty)$ become zero by following Röhrlich's method, then $dw/d\tau(\infty) = m_0\tau_0/c^2 \times g(dw/d\tau,dw/d\tau)w(\infty)$. We know only that the energy loss by radiation is finite by Eq.(36). The square of this equation is,

$$\frac{m_0}{\tau_0} \times \frac{m_0\tau_0}{c^2} g\left(\frac{dw}{d\tau},\frac{dw}{d\tau}\right)(\infty) = \left[\frac{m_0\tau_0}{c^2} g\left(\frac{dw}{d\tau},\frac{dw}{d\tau}\right)\right]^2(\infty). \tag{39}$$

Its solution is $m_0\tau_0/c^2 \times g(dw/d\tau,dw/d\tau)(\infty) = 0$, since $g(dw/d\tau,dw/d\tau) \leq 0$. Therefore,

$$\lim_{\tau \to 0}\frac{dw}{d\tau} = 0. \tag{40}$$

Therefore, my equation (33) can satisfy Röhrlich's stability condition. The stability of Eq.(33) was demonstrated in a two-stage analysis.

### 3.3   Calculations

As the final section in this section, I will present numerical calculation results showing the behavior of each model in a laser-electron interaction. The models are Eq.(33), the SZK equation Eq.(6) and the Landau-Lifshitz (LL) equation which is the major method for simulations. The form of the LL equation is as follows [18]:

$$m_0\frac{dw}{d\tau} = -eF_{\text{ex}}\left[\bullet, w^\flat + \frac{\tau_0}{m_0}f_{\text{ex}}^\flat\right] - e\tau_0\frac{dF_{\text{ex}}}{d\tau}[\bullet, w^\flat] + \frac{\tau_0}{m_0c^2}g(f_{\text{ex}},f_{\text{ex}})w \tag{41}$$

I chose the parameters of the Extreme Light Infrastructure - Nuclear Physics (ELI-NP) for calculations [19, 20]. The characteristic point of Eq.(33) is the term $-e(\eta g_0)c\mathbf{B}_{\text{ex}}$ which is derived from $-e\eta g_0 * \mathfrak{F}[\bullet, w^\flat]$. Therefore, we need to consider the condition in which an electron feels this force strongly. It will be the electron injection along $\mathbf{B}_{\text{ex}}$ field [Fig.2].

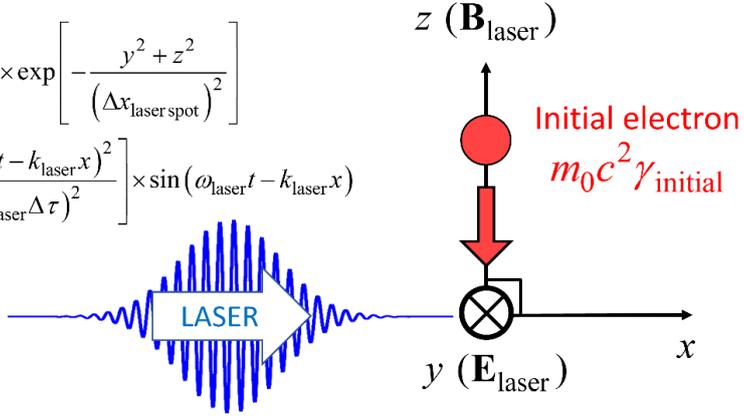

Fig.2    Setup of laser - electron "90 degree collision".

The laser propagates along the $x$ axis with the intensity of $1\times10^{22}\,\text{W}/\text{cm}^2$, the pulse width of 22fsec and the laser wavelength of $0.82\mu\text{m}$. An electron travels in the negative $z$ direction, which is the direction of the $\mathbf{B}_{\text{laser}}$ field. The energy of the electron is 700MeV.

The peak intensity of the laser is $1\times10^{22}\,\text{W}/\text{cm}^2$ which is a Gaussian shaped plane-wave like Eq.(28,29) in reference [12]. The pulse width is 22fsec and the laser wavelength is $0.82\mu\text{m}$. The electric field is set in the $y$ direction, the magnetic field is in the $z$ direction. The electron travels in the negative $z$ direction, with the energy of 700MeV initially. The numerical calculations were carried out by using the equations in the laboratory frame.

The radiation reaction appears directly in the time evolution of the electron's energy. I show this in Fig.3. The energy drop refers to radiation energy loss of the electron.   This figure is the best figure for understanding the behavior of radiation reaction. We can say that the solutions have almost the same tendency to very high accuracy. In particular, Eq.(33) and the SZK equation overlap completely. Therefore they can't be distinguished in this figure and from the final energy of the electron. The final energy of Eq.(33) and SZK equation are 302.8MeV and the LL equation is 301.1MeV, the energy difference is $O(1\text{MeV})$. The explanation of the convergence between the SZK and the LL equation is in reference [12]. Therefore, I will present the convergence between Eq.(33) and the SZK equation.

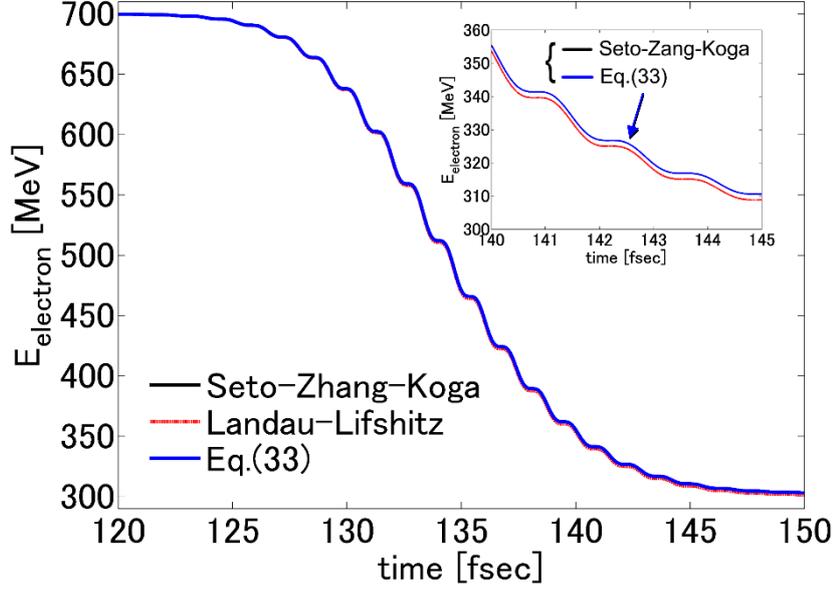

Fig.3  The energy of the electron. All of the models converged.
The final electron's energies are, Seto-Zhang-Koga: 302.8MeV, the Landau-Lifshitz: 301.1MeV and Eq.(34) 302.8MeV. The inset is a close-up of the figure.

The key parameters are $\eta f_0$ and $\eta g_0$. Their plots are shown in Fig.4. From these figures, we found that the order of them are $\eta f_0 = \mathrm{O}(10^{-8})$ and $\eta g_0 = \mathrm{O}(10^{-10})$. I introduced the term $-e\eta g_0 * \mathfrak{F}[\bullet, w^\flat]$ as the feature of Eq.(33). In the rest frame, this term becomes $|-e(\eta g_0)c\mathbf{B}_{\mathrm{ex}}|_{\mathrm{rest}} \sim 10^{-10} \times |-e\mathbf{E}_{\mathrm{ex}}|_{\mathrm{rest}}$. Thus, this new term is rounded into the external field like $-e\mathfrak{F}[\bullet, w^\flat] - e\eta g_0 * \mathfrak{F}[\bullet, w^\flat] \sim -e\mathfrak{F}[\bullet, w^\flat]$. For these reasons, Eq.(33) transforms to the SZK equation (6) as follows:

$$m_0 \frac{dw}{d\tau} = -\frac{e}{1-\eta f_0}\{\mathfrak{F}[\bullet, w^\flat] + \eta g_0 * \mathfrak{F}[\bullet, w^\flat]\}$$

$$= -\frac{e}{1-\eta\langle F_{\mathrm{LAD}} | F_{\mathrm{LAD}}\rangle}\mathfrak{F}[\bullet, w^\flat] + \mathrm{O}\bigl(\eta\langle F_{\mathrm{LAD}} | F_{\mathrm{ex}}\rangle, \eta\langle F_{\mathrm{LAD}} |*F_{\mathrm{ex}}\rangle\bigr) \quad (42)$$

Here, $\mathrm{O}\bigl(\eta\langle F_{\mathrm{LAD}} | F_{\mathrm{ex}}\rangle\bigr) = \mathrm{O}\bigl(\eta\langle F_{\mathrm{LAD}} |*F_{\mathrm{ex}}\rangle\bigr) = \mathrm{O}(\eta g_0)$ since the external field satisfies $\langle F_{\mathrm{ex}} | F_{\mathrm{ex}}\rangle = 0$ and $\langle F_{\mathrm{ex}} |*F_{\mathrm{ex}}\rangle = 0$. My new equation (33) as the extension from our previous SZK equation has good properties for numerical calculations. We can say the method using the LL equation, which is the first order perturbation of the LAD, is nearly equal to the suppression due to the effects of quantum vacuum.

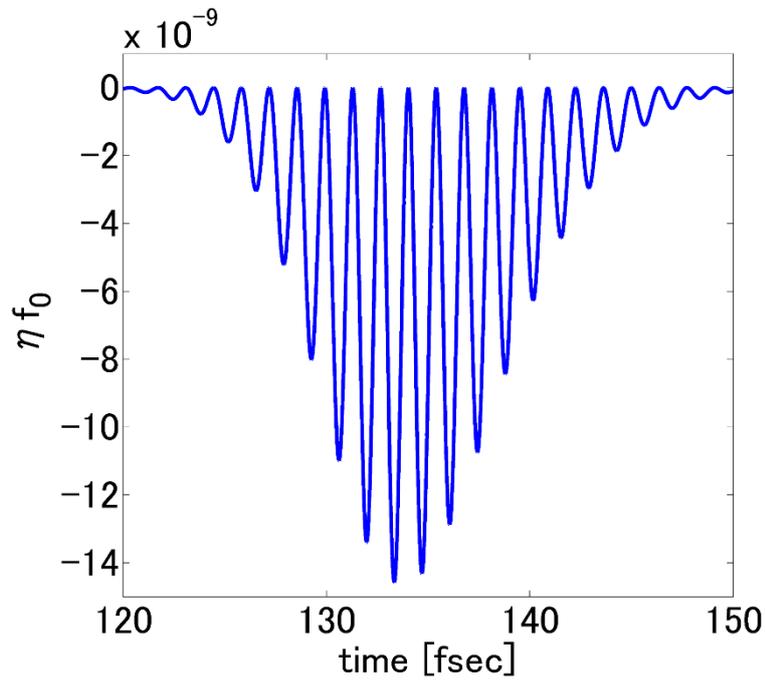

(a)   $\eta f_0$

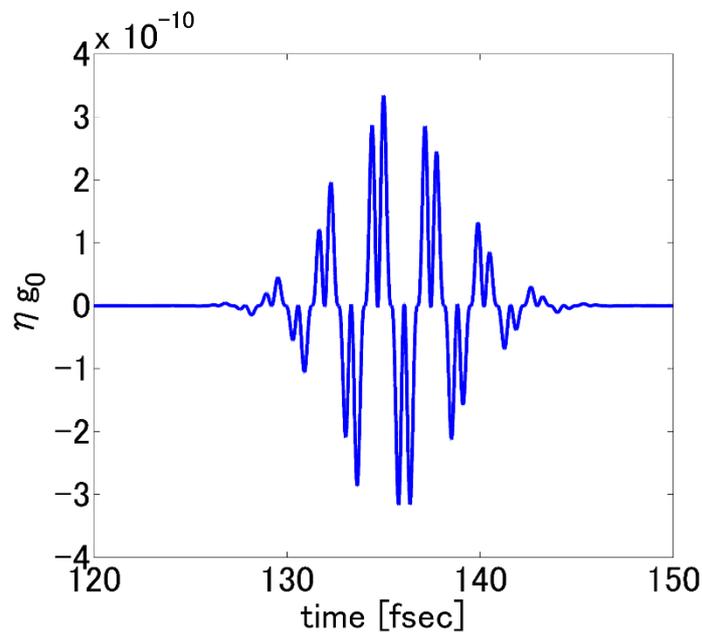

(b)   $\eta g_0$

Fig.4    Time evolution of factors   $\eta f_0$   and   $\eta g_0$

**4. Conclusion**

In summary, I updated our previous equation of motion with radiation reaction in quantum vacuum, naturally. The idea of the derivation of the new equation is same as our previous paper [12], however, the biggest difference is the introduction of the external field effects by following replacement [Eq.(12-13)].

$$F - \eta f \times F - \eta g \times {}^*F = F_{\text{LAD}} \quad \Rightarrow \quad F - \eta f \times F - \eta g \times {}^*F = F_{\text{ex}} + F_{\text{LAD}} \tag{43}$$

Via this replacement, the new model includes the radiation-external field's interactions. Now we rewrite Eq.(25) as,

$$\boxed{\frac{dw}{d\tau} = -\frac{e}{m_0}\mathcal{K}[\bullet, w^\flat, \mathfrak{F}^\flat] \quad \in \mathbb{V}_M^{\ 4}} \tag{44}$$

or

$$\frac{dw^\mu}{d\tau} = -\frac{e}{m_0(1-\eta f_0)}\left(\mathfrak{F}^{\mu\nu} + \eta g_0 {}^*\mathfrak{F}^{\mu\nu}\right)w_\nu. \tag{45}$$

This equation is the important result in this paper. In theoretical analysis, I was able to achieve the avoidance of the run-away in Heisenberg-Euler vacuum under Eq.(33), $1 - 2\eta \mathbf{E}_{\text{ex}}^{\ 2}|_{\text{rest}}/c^2 > 0$. From the results of numerical calculation, it showed that Eq.(45) agrees well with the LL equation (41). It represents the first order perturbation of the LAD equation is nearly equivalent to the run-away suppression by quantum vacuum. Here, I focus on the tensor of $e/m_0 \times \mathcal{K} \in \mathbb{V}_M^{\ 4} \otimes \mathbb{V}_M^{\ 4} \otimes \mathbb{V}_M^{\ 4} \otimes \mathbb{V}_M^{\ 4}$. This is the generalization of our previous charge to mass ratio [12],

$$\frac{Q}{M} = \frac{e}{m_0(1-\eta\langle F_{\text{LAD}} | F_{\text{LAD}}\rangle)} = \frac{e}{m_0} + \frac{\delta e}{m_0} \in \mathbb{R}. \tag{46}$$

The charge-mass particle system is built on measure theory. Now the mass measure is denoted by $m : \mathbb{A}^4 \to \mathbb{R}$ and the general charge measure including anisotropy is defined as the tensor function $\mathcal{E} : \mathbb{A}^4 \to \mathbb{V}_M^{\ 4} \otimes \mathbb{V}_M^{\ 4} \otimes \mathbb{V}_M^{\ 4} \otimes \mathbb{V}_M^{\ 4}$ in Minkowski spacetime. The equation of motion should be described as,

$$\boxed{dm(x)\frac{dw}{d\tau} = -d\mathcal{E}(x)[\bullet, w^\flat, \mathfrak{F}^\flat] \quad \in \mathbb{V}_M^{\ 4}}. \tag{47}$$

Since I considered a classical point particle, it will be based on the Dirac measure. But I will not consider the concrete form of the measures. This includes the unknown information on how the mass and charge themselves are described. However, the relation between $dm(x)$ and $d\mathcal{E}^{\mu\nu\alpha\beta}(x)$ is very important. The measure can be connected to others via the derivative like $d\mathcal{E}^{\mu\nu\alpha\beta} = (d\mathcal{E}^{\mu\nu\alpha\beta}/dm)\,dm$. This $d\mathcal{E}^{\mu\nu\alpha\beta}/dm$ is called the Radon-Nikodym derivative [21].

$$dm(x)\left(\frac{dw}{d\tau}+\frac{d\mathcal{E}}{dm}[\bullet,w^{\flat},\mathfrak{F}^{\flat}]\right)=0 \quad \Rightarrow \quad \frac{dw}{d\tau}+\frac{d\mathcal{E}}{dm}[\bullet,w^{\flat},\mathfrak{F}^{\flat}]=0 \qquad (48)$$

This equation must become equivalent to Eq.(44). Therefore, the Radon-Nikodym derivative becomes,

$$\boxed{\frac{d\mathcal{E}}{dm}=\frac{e}{m_0}\mathcal{K}=\frac{e}{m_0}\frac{G+\eta g_0\times\frac{\varepsilon}{2!}}{1-\eta f_0}} \quad \in \mathbb{V}_M^{\ 4}\otimes\mathbb{V}_M^{\ 4}\otimes\mathbb{V}_M^{\ 4}\otimes\mathbb{V}_M^{\ 4} \qquad (49)$$

Here, $G=g^{\mu\alpha}g^{\nu\beta}\partial_\mu\otimes\partial_\nu\otimes\partial_\alpha\otimes\partial_\beta$. This is a generalization of the charge to mass ratio by Fletcher and Millikan [15, 16] including the anisotropy of quantum vacuum.

**Acknowledgements**

I thank Dr. James K. Koga and Dr. Sen Zhang for discussion. This work is supported by Extreme Light Infrastructure – Nuclear Physics (ELI-NP) – Phase I, a project co-financed by the Romania Government and European Union through the European Regional Development Fund, and also partly supported under the auspices of the Japanese Ministry of Education, Culture, Sports, Science and Technology (MEXT) project on "Promotion of relativistic nuclear physics with ultra-intense laser."